\documentclass{IEEEtran}
\IEEEoverridecommandlockouts
\usepackage{cite}
\usepackage{amsmath,amssymb,amsfonts}
\usepackage{algorithmic}
\usepackage{graphicx}
\usepackage{textcomp}   
\usepackage{xcolor}
\usepackage{multirow}
\usepackage{hyperref}
\usepackage{caption}
\usepackage{subcaption}
\usepackage[export]{adjustbox}
\usepackage{comment}
\def\BibTeX{{\rm B\kern-.05em{\sc i\kern-.025em b}\kern-.08em
    T\kern-.1667em\lower.7ex\hbox{E}\kern-.125emX}}
    
\makeatletter
\newcommand{\linebreakand}{
  \end{@IEEEauthorhalign}
  \hfill\mbox{}\par
  \mbox{}\hfill\begin{@IEEEauthorhalign}
}
\makeatother        
\newcommand{\myceil}[1]{\left \lceil #1 \right \rceil }
\usepackage{etoolbox}
\makeatletter
\patchcmd{\@makecaption}
  {\scshape}
  {}
  {}
  {}
\makeatletter
\patchcmd{\@makecaption}
  {\\}
  {.\ }
  {}
  {}
\makeatother
    
\def\tablename{Table}

\begin{document}

%\title{Combining Evidence from Vocal Tract and Voice Source Features for Synthetic Speech Classification
%%% Alternative %%%
\title{Evince the artifacts of Spoof Speech by blending Vocal Tract and Voice Source Features
% \thanks{All the authors are with Speech Information Processing (SIP) Lab, Department of Electrical Engineering, Indian Institute of Technology Hyderabad, Kandi, Telangana 502284, India.\\
% }
}

\author{
Tadipatri Uday Kiran Reddy ,
Sahukari Chaitanya Varun,
Kota Pranav Kumar\\
Sankala Sreekanth,
Kodukula Sri Rama Murty\\
Indian Institute of Technology Hyderabad, Telangana, India\\
\textit{\{ee19btech11038, ee19bech11040, ee19btech11051, ee20resch11011, ksrm\} @iith.ac.in}
}

% The paper headers
\markboth{This work will be submitted to the IEEE for possible publication.}%
{}

\maketitle

\begin{abstract}
With the rapid advancement in synthetic speech generation technologies, great interest in differentiating spoof speech from the natural speech is emerging in the research community. The identification of these synthetic signals is a difficult task not only for the cutting-edge classification models but also for humans themselves. To prevent potential adverse effects, it becomes crucial to detect spoof signals. From a forensics perspective, it is also important to predict the algorithm which generated them to identify the forger. This needs an understanding of the underlying attributes of spoof signals which serve as a signature for the synthesizer.
This study emphasizes the segments of speech signals critical in identifying their authenticity by utilizing the Vocal Tract System(\textit{VTS}) and Voice Source(\textit{VS}) features.

In this paper, we propose a system that detects spoof signals as well as identifies the corresponding speech-generating algorithm. We achieve 99.58\% in algorithm classification accuracy. From experiments, we found that a VS feature-based system gives more attention to the transition of phonemes, while, a VTS feature-based system gives more attention to stationary segments of speech signals. We perform model fusion techniques on the VS-based and VTS-based systems to exploit the complementary information to develop a robust classifier. Upon analyzing the confusion plots we found that WaveRNN is poorly classified depicting more naturalness. On the other hand, we identified that synthesizer like Waveform Concatenation, and Neural Source Filter is classified with the highest accuracy. Practical implications of this work can aid researchers from both forensics (leverage artifacts) and the speech communities (mitigate artifacts).

\end{abstract}


%%%%%%%%%%%%%%%%%%%%%%%%%%%%%%%%%%%%%%%%%%%%%%%%%%%%%%%%%%%%%%%%%%%%%%%%%%%%%%%%%%%%%%%%%%%%%%%%%%%%%%%%%%%%%%%%%%%%
% \input{intro}
\section{Introduction}

Speech synthesis is the process of generating natural speech using machines. With the rapid improvements in deep learning technologies, speech synthesis algorithms have remarkably improved the genuineness of synthesized speech. The advances in speech synthesis algorithms made them deploy in various applications - aiding vocally handicapped, assisting vision-impaired humans through screen readers, voice dubbing in multi-media applications, and human-machine interactions, etc. Although speech synthesis algorithms have abundant usage in improving the quality of life, they pose a severe threat in forensic science if used negatively. For example, the advancements in speech synthesis algorithms severely threaten the usage of biometric voice systems—especially the voice biometrics deployed in machine-critical areas such as financial transactions and forensic applications. The forger may exploit the speech synthesis systems to create illegal audio copies to defame public figures. The recent occurrence of mishaps \cite{fraudcase_1,fraudcase_2} illustrates the adverse consequences caused by voice impersonators. Following the threats of synthesis systems, it is vital to study the defence strategies that develop an algorithm to discriminate synthetic speech from natural speech. Correspondingly, there have also been many studies that investigate the usage of different features that could be used for differentiating spoof signals from natural signals. Some studies suggest that features consisting of high-frequency information have pronounced effects on spoof speech \cite{high_freq_artifacts, feature_comp_synth_detect}. \\
Although the recent models perform reasonably well in discriminating synthetic speech from natural speech, they lack an interpretation of the artifacts captured by the model. Finding the artifacts captured by the model to detect the fake speech will aid forensic experts in narrowing down their search area for finding the synthesizer. Detecting the underlying synthesizer helps identify the falsifier. It also helps reduce models' complexity by explicitly capturing the possible artifacts in the feature extraction. Moreover, the synthetic speech groups could focus more on them to improve the naturalness of the generated speech. 

Naturally, speech is produced by modulating the expelled air from the lungs through the movements of articulators (nose, mouth, etc.). Hence the air from the lungs acts as a source of energy, and articulators act as a modulator to produce intelligible sounds called phonemes. Natural variations are inevitable in delivering speech sounds. Majorly the natural variations are introduced at the system and source levels while generating the speech. Speech synthesis algorithms aim to mimic this natural system and may fail to synthesize the subtle variations introduced by the human speech production system.

In that direction, the authors of \cite{dnn_deepfake_detect} claimed that there exist certain phonemes that synthesized models are not able to generate precisely. They empirically showed that synthetic systems fail to generate fricatives (/S/,/SH/), nasals (/M/,/N/), and vowels (/Y/) compared to the other phonemes. Similarly, the work in \cite{silence_artifacts} shows that the silence regions contain discriminative patterns to classify synthetic speech from natural speech. They found that synthetic models face difficulty in synthesizing the statistically-realistic silence intervals. 

In the literature, various features are used to capture the artifacts introduced at the Vocal Track System (VTS) and Voice Source (VS). Spectral magnitude features (Log-Mel filter bank energies) are mainly used to capture VTS characteristics. To analyze VS, Linear Prediction (LP) residuals are used extensively. Many works \cite{lp_speaker_recog_1,lp_speaker_recog_2,lp_speaker_recog_3,lp_speaker_ver,lp_speaker_info} have shown the complementary information present on these features. Hence, to capture and improve the performance of synthetic systems, it is essential to explore both features.

This work explores spectral magnitude features ( Log-Mel filter bank energies ) and LP residuals to perform synthetic speech algorithm classification. As the spectrum of LP residuals is relatively flat, we have proposed an end-to-end feature extraction module to extract more discriminative patterns from the LP residuals. The spectral magnitude features from the speech signal or the end-to-end features learned from the LP residual are used with the X-Vector \cite{xvect} framework to perform synthetic speech classification. Further, the attention layer in the X-Vector architecture is used to analyze the artifacts learned by the model in discriminating the synthetic speech. We found that the synthetic speech detection systems trained with different features tend to attend to various artifacts of synthesis algorithms. Specifically, the VTS features tend to attend more to stationary speech segments. On the other hand, the VS features tend to attend more to the co-articulation variations, i.e., phoneme transitions. Later, the subsystems built on different features are fused at different architecture levels to improve classification accuracy. Finally, we identify two sets of algorithms that perform worst using our system, indicating more naturalness and, on the contrary, a few which are perfectly classified, i.e., more synthetic. Implications of this work can aid researchers from forensics (leveraging artifacts) and speech synthesis (mitigating the artifacts). \\

The rest of the paper is organized as follows. Section $\mathrm{II}$ describes the source-system model of the speech production mechanism and motivates the need to extract VTS features and VS features. Section $\mathrm{III}$ describes the details of the VTS feature extraction. Section $\mathrm{IV}$ describes the DNN architecture of a synthetic speech classifier. Section $\mathrm{V}$ describes the VS feature extraction and DNN models to extract high-level representations from it. Subsequently, Section $\mathrm{VI}$ contains information about the experimental setup, including the implementation details and datasets used. Section $\mathrm{VII}$ discusses the results of the models used and includes details of the experiments conducted by fusion. In addition, it also contains a comparison of the proposed models and their variants. Section $\mathrm{VII}$ analyzes the artifacts learned by the synthetic speech classification models by exploring the attention modules and proposes a method to fuse the subsystems to improve the performance. Finally, $\mathrm{VIII}$ concludes the paper with a summary of contributions and the future scope of the work.

%%%%%%%%%%%%%%%%%%%%%%%%%%%%%%%%%%%%%%%%%%%%%%%%%%%%%%%%%%%%%%%%%%%%%%%%%%%%%%%%%%%%%%%%%%%%%%

\section{Motivation for feature extraction}
In this section, we briefly describe one of the speech production models known as the Source-System model.

In the literature, many researchers have proposed models \cite{speech_prod_complete, arx_model,dynamic_model, GA_model,models_survey} to describe the speech production system and model the non-linearities. However, in this study, we restrict ourselves to Source-System Model as described in the paper \cite{source_filter}. This model assumes linearity and that the source and system work independently to produce speech.

\begin{gather}
\label{eq:speech_conv}
    s[n] = v[n]*e[n]
\end{gather}

Where $s[n]$, $v[n]$, and $e[n]$ are the speech signal, impulse response of the vocal tract filter, and excitation signal. The speech signal is the output of the time-varying digital filter ($v[n]$) with time-varying excitation ($e[n]$) \cite{anthropometry_voice}, which can be mathematically modeled as Eq \ref{eq:speech_conv}.

\begin{figure}[h!]
    \centering
    \includegraphics[width=\linewidth]{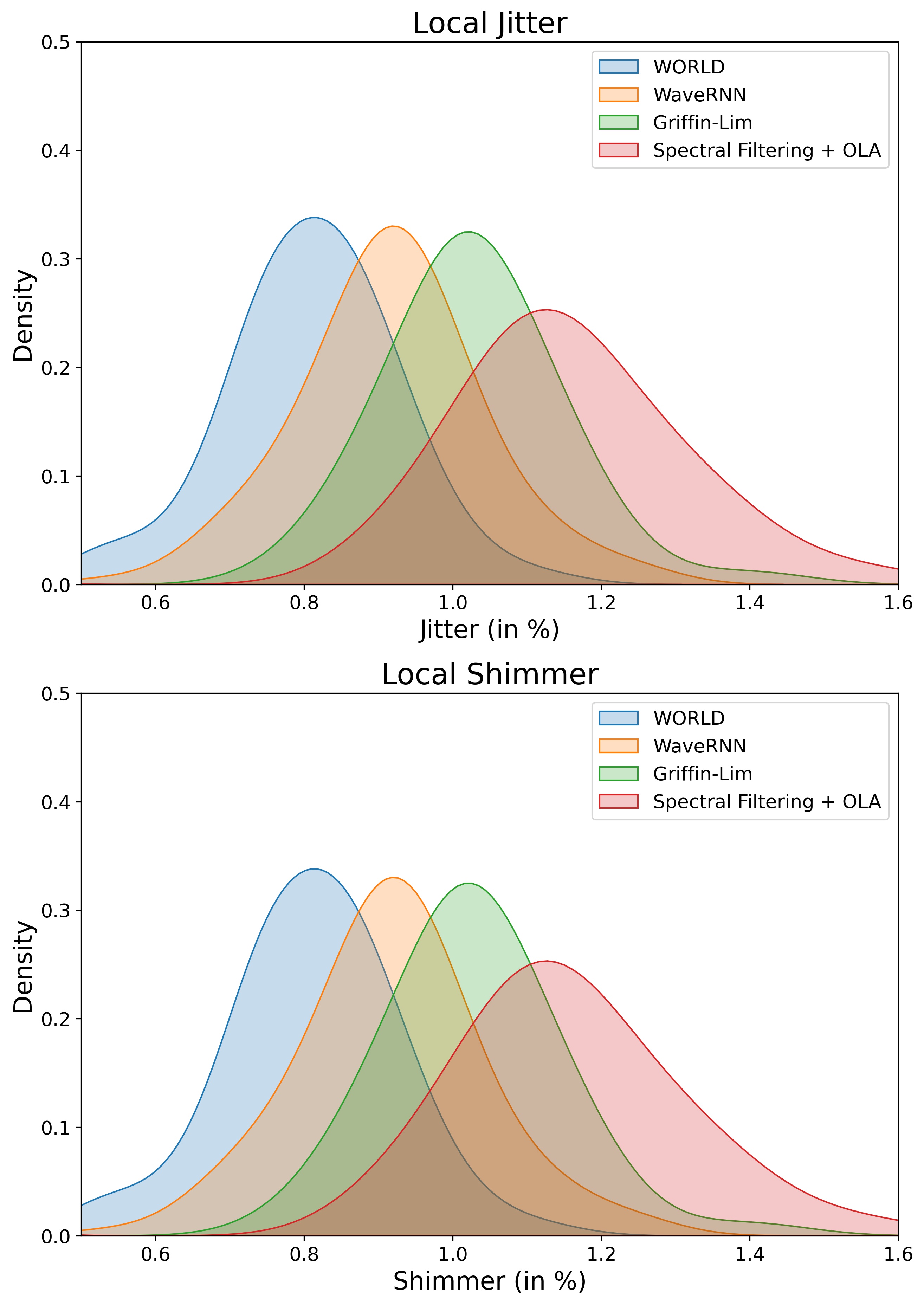}
    \caption{Histogram of Local Jitter and Local Shimmer.}
    \label{fig:shim_jit}
\end{figure}    

\begin{figure*}[ht!]
	\centering
	\begin{minipage}{1.0\columnwidth}
		\centering
		\centerline{\includegraphics[width=0.88\textwidth]{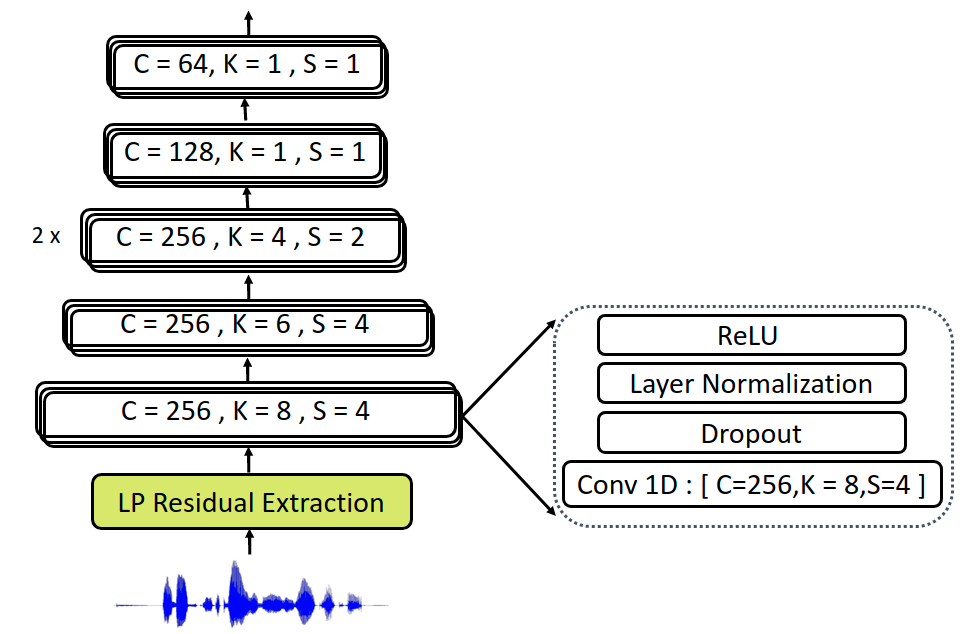}}
		\caption{Feature extraction from linear prediction residuals}
		\label{fig:arch3}
	\end{minipage}%
	\begin{minipage}{1.0\columnwidth}
		\centering
		\centerline{\includegraphics[width=0.99\textwidth]{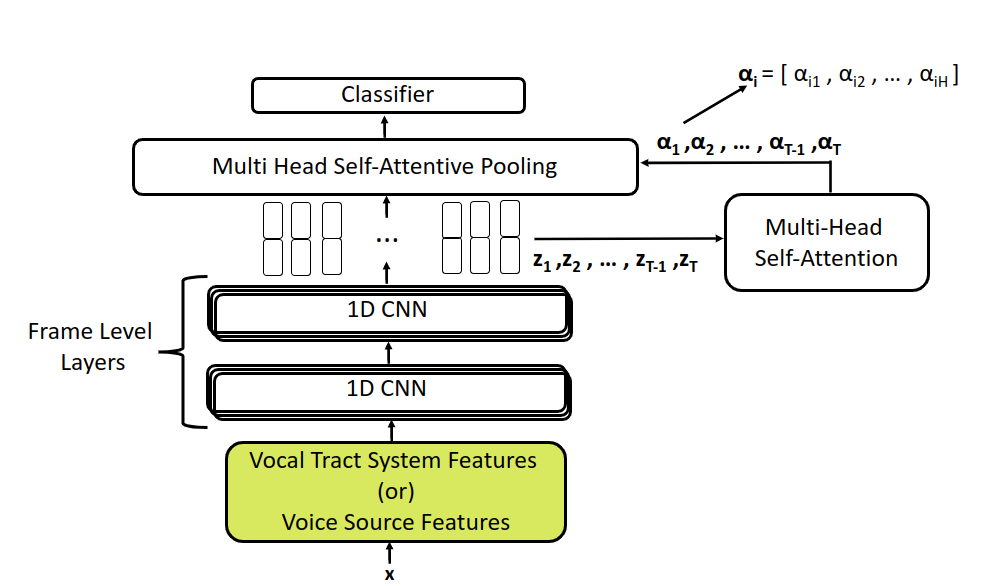}}
		\caption{Synthetic speech classification architecture}
		\label{fig:arch2}
	\end{minipage}
\end{figure*}

Synthetic speech algorithms seek to mimic this natural system, where they may fail to incorporate subtle features of speech production, such as
\begin{itemize}
    \item \textit{Prosodic variations such as jitter, shimmer}:
    
    \textit{Jitter} is a parameter to capture frequency variations in a time period as described in Eq \ref{eq:jitt}. Here $T_i$ is the duration of a period and $N$ is the number of periods.
    \begin{equation}
        \label{eq:jitt}
        \text{Jitter (in \%)} = 100\frac{N}{N-1}\frac{\sum_{i=1}^{N-1}|T_i - T_{i+1}|}{\sum_{i=1}^{N}T_i}
    \end{equation}
    \textit{Shimmer} is a parameter to capture amplitude variations in a time period as described in Eq \ref{eq:shim}. Here $A_i$ is the amplitude in period and $N$ is the number of periods.\newline
    \begin{equation}
        \label{eq:shim}
        \text{Shimmer (in \%)} = 100\frac{N}{N-1}\frac{\sum_{i=1}^{N-1}|A_i - A_{i+1}|}{\sum_{i=1}^{N}A_i}
    \end{equation}
    The excitation source of synthetic speech is more periodic and less noisy; therefore less jitter and shimmer in the impulse train, as pointed out in the paper \cite{jitter_spoof}. Among synthetic speech algorithms, there could be distinct patterns in these impulse trains, causing variations in jitter and shimmer. Figure \ref{fig:shim_jit} shows the histogram of Local Jitter and Local Shimmer of different synthetic speech signals. It is evident that jitter and shimmer variations among these algorithms are different, which gives us the motivation to look at the features that capture such information.
    \item \textit{Co-articulation of phoneme sequences}:
    
    {Coarticulation is the context-dependent variation of phonemes' articulatory and acoustic realization, especially of consonants. The generation of highly natural and intelligible speech requires the simulation of realistic articulatory movements. Phoneme transitions in synthetic speeches tend to be sharper than that observed in natural peech\cite{coarticulation, story_2017, birkholz_2013}}. Hence, we could emphasize the coarticulation capturing features for the classification problem to attribute as discriminatory evidence for the given synthesizer.
\end{itemize}

In the following sections, we discuss the features which could potentially capture the artifacts in spoof speech; Log Mel Spectrogram, capturing \textit{VTS} features, and Linear Predictive Residuals, capturing \textit{VS} features.

\section{Vocal Tract System Features: Log-Mel filter bank energies}

The Vocal Tract System can be considered a concatenation of acoustic tubes. The resonances of the vocal tract manifest as formant frequencies in the spectrogram of the speech signal. Hence capturing the formant frequencies of the speech signal can convey information about the vocal tract system used in the speech production process. Many works \cite{MFCC_alzheimers,env_sound_classify,poser_1990,acoustic_event_detect,auditory_scene_recog} utilize the \textit{MFCC} (mel-frequency cepstral coefficients), and \textit{LPCC} (linear prediction cepstral coefficients) features to capture the magnitude information. 

Log Mel filter bank energies are also one of the popular ways to extract information about the vocal tract system\cite{logmel_asfeature}. The extracted energies consist of up to 2nd-order statistics conveying vocal tract features in the speech production model.

\section{Synthetic speech classification - X-vector }

This section provides the details of the synthetic speech classification architecture. X-vector framework \cite{xvect} originally proposed for speaker classification is explored to classify synthetic speech from natural speech. As shown in Figure \ref{fig:arch2}, the X-vector framework consists of three submodules: (i) Frame level encoder, (ii) Multi-head self-attentive pooling layer, and (iii) segment level encoder. Frame level encoder comprises 1D convolutional layers and is framed to extract high-level features representing the artifacts introduced by the synthetic algorithms. Multi-head self-attentive pooling layer attentively pools the information from variable-length high-level representations into a fixed dimensional embedding. Finally, a set of Dense layers in the segment-level encoder outputs the logits of the classes. Finally, the cross-entropy loss function is used to train the network.

\subsection{Multi-Head Self-Attention}

In this work, we assume that all the frames may not be equally important to discriminate the synthetic algorithms. The results in \cite{dnn_deepfake_detect} validate the assumption by showing certain phonemes (/S/,/SH/,/M/,/N/) are relatively more important than other phonemes. Hence we have used multi-head self-attention modules to highlight the speech regions that are more discriminated among the synthetic algorithms. In further sections, we identify the artifacts of the synthetic algorithms by plotting these attention values over the speech signal, as shown in Figure \ref{fig:atten}.

Let $(\textbf{x}_{i})_{i=1}^{T} \in \mathbb{R}^n$ be the sequence of feature vectors extracted from the input speech signal and $f:\mathbb{R}^{n \times T} \mapsto \mathbb{R}^{d \times T'}$ be the frame level encoder function in the X-vector framework. Frame-level encoder extracts high-level representations from the sequence of features vectors
\begin{equation}
    \textbf{h}_{1},\textbf{h}_{2},...,\textbf{h}_{T'} = f(\textbf{x}_{1},\textbf{x}_{2},...,\textbf{x}_{T_{i}})
\end{equation}
Where $(\textbf{h}_{i})_{i=1}^{T'} \in \mathbb{R}^{d}$ are the high-level representations extracted from the frame level encoder. Time axis dimension $T'$ depends on the effective receptive field, and stride \cite{receptive_field} of the frame level encoder $f$.

Self-attention module relatively weights the frame level representations and computes weighted stats (mean, variance) to obtain the fixed dimensional embedding. The relative importance of the frame-level representations is calculated using the encoder-decoder framework. The encoder in this work consists of the 1D convolutional layer and projects the frame-level representations to lower dimensional space. The decoder calculates the relative importance of frames using the encoded information. Let $s_{e}:\mathbb{R}^{d \times T'} \mapsto \mathbb{R}^{b \times T'}$ and $s_{d}:\mathbb{R}^{b \times T'} \mapsto \mathbb{R}^{d \times T'}$ be the encoder and decoder in the self-attention module. Then the relative importance is computed as follows 
\begin{equation}
    \textbf{e}_{1}, \textbf{e}_{2},...,\textbf{e}_{T'} = s_{e}(\textbf{h}_{1},\textbf{h}_{2},...,\textbf{h}_{T}')
\end{equation}
\begin{equation*}
    \textbf{a}_{1}, \textbf{a}_{2},...,\textbf{a}_{T'} = Softmax(s_{d}(\textbf{e}_{1}, \textbf{e}_{2},...,\textbf{e}_{T'}))
\end{equation*}

Where $(\textbf{e}_{i})_{i=1}^{T'} \in \mathbb{R}^b$ are low dimensional encoded frames, $(\textbf{a}_{i})_{i=1}^{T'} \in \mathbb{R}^d$ are relative weights of $(\textbf{h}_{i})_{i=1}^{T} \in \mathbb{R}^d$ computed from the self-attention module and $Softmax(.)$ activation function is operated along the time axis. Notice that the self-attention module assigns the relative weights to both time and channel dimensions. Further the statistics of the relatively weighted frame-level representation ( $(\textbf{a}_{i} \odot \textbf{h}_{i})_{i=1}^{T}$ ) are computed to obtain the fixed dimensional embedding. A single self-attention layer may not be able to efficiently assign relative weights to multiple sets of complimentary feature vectors. Hence, the multi-head self-attention module incorporates multiple self-attention layers to obtain the fixed dimension embeddings to capture various aspects of frame-level representations efficiently. The fixed-dimensional embeddings extracted from each head of the multi-head self-attentive module are concatenated to incorporate the information from multiple views.  

\begin{table}[h!]
\centering
\caption{Effective receptive field and stride of proposed models}
\begin{tabular}{|l|l|l|}
\hline
\textbf{Architecture} & \textbf{Receptive Field} & \textbf{Effective Stride} \\ \hline
LPR-DNN & 2160 (135 ms) & 64 (4 ms) \\ \hline
LMS-DNN & 4960 (310 ms) & 12 (0.75 ms) \\ \hline
\end{tabular}
\label{tab:receptive_field}
\end{table}

\subsection{LMS-DNN: VTS feature-based DNN}
To exploit the vocal tract features, we built an X-Vector \cite{xvect} network using Log-Mel Spectrogram features as shown in Figure \ref{fig:arch2}. The Mel Spectrogram features are passed into frame level encoder. After obtaining a compact representation $(128 \times T')$, $T'$ is governed by the receptive field and stride as shown in Table \ref{tab:receptive_field}. Next, we perform a self-attention module ($s_e: \mathbb{R}^{128\times T'} \mapsto \mathbb{R}^{64\times T'}$, and $s_d: \mathbb{R}^{64\times T'} \mapsto \mathbb{R}^{128\times T'}$) followed by a pooling layer to normalize the features across time steps to obtain a fixed 256 (concatenation of mean and variance of a 128-dimensional vector)-dimensional vector. Finally, we use the logistic regression layer to perform algorithm classification.

\begin{figure}[h!]
    \centering    
    \includegraphics[width=0.45\textwidth]{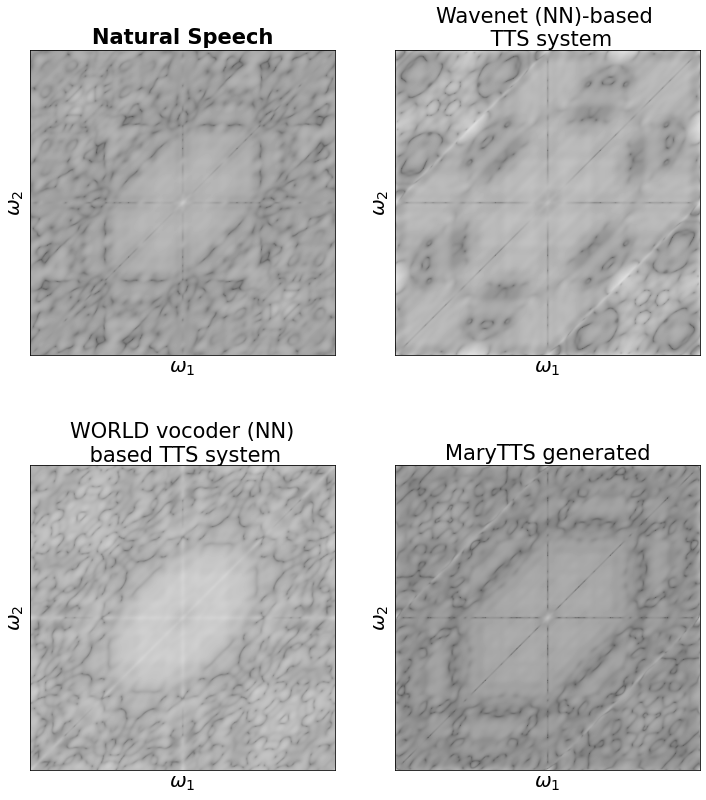} 
    \caption{Log-applied bicoherence magnitude plots on LP Residuals for samples of 1 natural and 3 different synthetic speech algorithms with common speaker (\textit{LA\_0089}) and common transcript (\textit{IS THERE ON HIS HANDS.}). These plots are displayed on a common intensity scale of [10,25] to enhance visibility. }
    \label{fig:bicoher}
\end{figure}

\section{Voice Source Features: Linear Prediction Residuals}
\label{section:lpres_section}
Capturing the magnitude spectral features (2nd-order statistics) for speaker verification\cite{spkrverif} tasks has been one of the standard approaches. The DL community has developed TTS systems that primarily rely on loss functions that minimize the distance between 2nd Order statistics \cite{tachetron, wavenet, deep_voice}.  Therefore, using only 2nd-order features might not provide complete discriminatory evidence among algorithms. This emerges the need for looking at features that capture higher-order statistics.

There have been some established works \cite{lp_speaker_recog_1,lp_speaker_recog_2,lp_speaker_recog_3,lp_speaker_ver,lp_speaker_info} which stated that the source and the system features, complementary within themselves, could be utilized together for analysis of speech signals. The voice source features to capture the vocal cords' characteristics in speech production. The phase spectral information is complementary to information from VTS features\cite{evidence_res_ph_mfcc,better_mfcc_feature} as we are limiting to 2nd-order statistics. And one such higher-order statistical feature exploited in this work is LP Residuals.

Linear Prediction (LP) analysis \cite{synth_uttr_roman,allpass_modelling_lpr, LP_tutorial} is a methodology where the magnitude spectral envelope is repressed by first finding the magnitude spectral envelope and then operated with an inverse formulation which results in residual rich in higher order correlations, with phase content. This feature, called the LP residual, captures the voice source characteristics. Analytically, the VTS response is modeled as a linear prediction filter. LP Residual is obtained as a result of inverse filtering the speech signal with an estimated VTS response.

\begin{gather}
    s[n] = \sum_{k=1}^{K}a_ks[n-k] + Ge[n] \\
    S(z) = V(z)E(z); \;\;\; V(z) = \frac{G}{A(z)} = \frac{G}{1 - \sum\limits_{k=1}^{K}a_{k}z^{-k}}\\
    E(z) = V(z)^{-1}S(z) 
\end{gather}
% \end{comment}

Where $V(z)$ is the VTS response, $a_k$ are the linear coefficients, and $K$ is the order of the Linear Prediction.

For our study, we perform all-pole modeling of LP Residual\cite{allpass_modelling_lpr}. The order is chosen such that we have one resonance for each kHz ($K = \myceil{4 + f_s/1000}$) \cite{rabiner_book}. We ensure that the order is odd to constrain the LP analysis with at least one real pole \cite{quatieri_book}.

{According to the Source-System model, the excitation source of voiced sounds is ideally an impulse train. However, natural signals are different from the ideal scenario, and the excitation source neither has a consistent period(jitter) nor a constant amplitude(shimmer), making it less predictable, as shown in Figure \ref{fig:shim_jit}. On the other hand, LP Residuals of spoof signals are periodic and have relatively constant periods making them more predictable, and this is consistent with all the spoof algorithms. This is a potential discriminative feature that can be exploited to classify spoof signals.}

To validate the use of this feature, we perform \textit{Bispectrum} \cite{cepstral_bispectral_analysis} analysis on the LP residuals, a third-order statistic (or skewness) study. We use \textit{bicoherence} (Equation \ref{eq : bicoher}), which is the normalized bispectrum making phase content more pronounced \cite{bicoher_phase}.

\begin{gather}
    B_{coher}(\omega _1, \omega _2) = \\ \nonumber
    \frac{1}{W}\frac{\sum_{w=0}^{W-1}X(\omega _1)X(\omega _2)X^*(\omega _1 + \omega _2)}{\sqrt{\sum_{w=0}^{W-1}|X(\omega _1)X(\omega _2)|^2\sum_{w=0}^{W-1}|X^*(\omega _1 + \omega _2)|^2}}
\label{eq : bicoher}
\end{gather}
As shown in Figure \ref{fig:bicoher}, the patterns observed in bicoherence plots\footnote{\url{https://in.mathworks.com/matlabcentral/fileexchange/3013-hosa-higher-order-spectral-analysis-toolbox}} suggest that the LP residuals have captured the higher-order correlations highlighting phase information (LP residuals are obtained from inversion of the magnitude spectrum of the signal) and support its utilization as discriminatory evidence. 

\begin{figure*}[h]
    \centering
    \includegraphics[width=1.75\columnwidth, height=15cm]{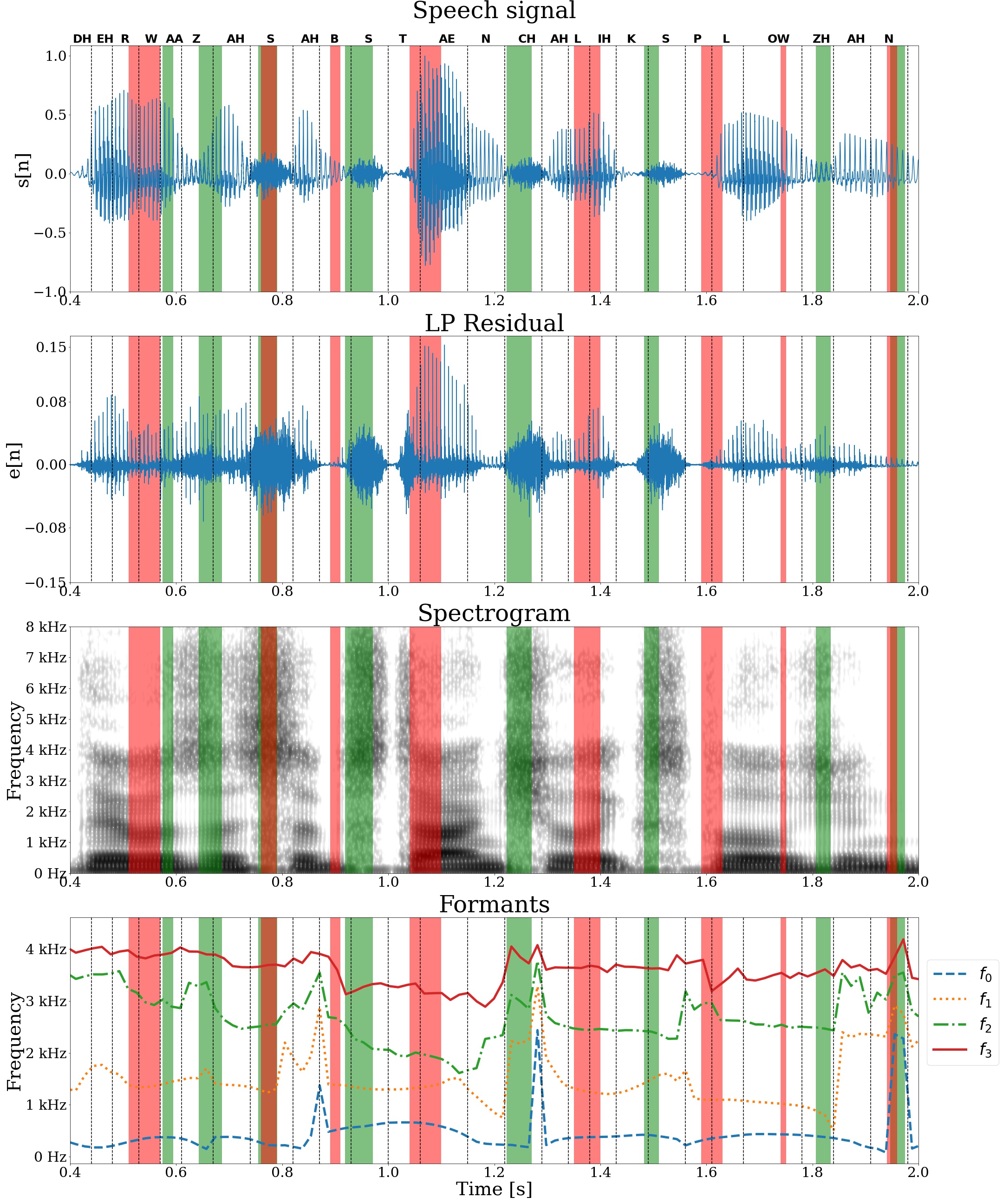}
    \caption{Attention maps on a synthetic speech signal with an utterance of "\textit{THERE WAS A SUBSTANTIAL EXPLOSION.}", NOTE:- GREEN bands indicate attention by LPR-DNN whereas RED represents LMS-DNN}
    \label{fig:atten}
\end{figure*}

Hence, we extracted more exhaustive information from a speech sample by fusing both features which are highlighted in Figure \ref{fig:atten}. 

\subsection{LPR-DNN: VS feature-based DNN}
% LP Residuals 
We extract low-dimensional features from the LP Residuals using 1D-convolutional filters and non-linear activation functions. LP Residuals are obtained by inverse filtering the speech signal with an envelope - the so-called VTS Response. Therefore the resultant spectrum of LP Residuals is relatively flat, and traditional feature extraction methods such as spectral magnitude will most likely be inefficient. This motivates us to introduce a feature extractor. Since the usage of LP Residuals for the Spoof detection/classification task is studied much less in the literature, we propose a DNN-based feature extractor consisting of 1D-Convolutional filter banks with non-linear activation functions. Figure \ref{fig:arch3} describes the parameters used in the feature extractor; we utilize $Dropout$ layers to equally train all the coefficients of 1D-filters \cite{dropout}. $Layer$ $Normalisation$ \cite{layer_norm} is applied at end of every layer to avoid vanishing gradients while training the network. We extract $(64 \times T)$ dimensional features from the feature extractor.

Later, frame-level encoding is performed on these features using the X-Vectors system. After obtaining a compact representation (128 x $T'$), where $T'$ is governed by receptive field and stride from Table \ref{tab:receptive_field}. The rest of the network follows the description provided in Figure \ref{fig:arch2}. It is worth noting that attention values will be different than that of LMS-DNN because here, inputs consist of additional information. Figure \ref{fig:atten} shows bands of attention on speech signal, and there is no overlap between LPR-DNN and LMS-DNN bands. This indicates the presence of complementariness in VTS features and VS features.  

\section{Experimental Setup}
\subsection{Implementation details}
The hardware specifications for conducting the studies are Intel Xeon Silver $4114$ CPU @ $2.20$GHz, and four NVIDIA GeForce RTX $2080$ Ti GPUs. The software packages utilized are \textit{PyTorch} library for training the neural networks and \textit{scikit} for computing LP Residuals. This link \footnote{\url{https://github.com/TUdayKiranReddy/SPCUP2022}}consists of source codes for replicating all the results reported in this paper.
% \begin{table*}[]
% \centering
% \begin{tabular}{|c|c|c|}
% \hline
% \textbf{ASVSpoof class} & \textbf{Input/Ouput} & \textbf{Generator} \\ \hline
% A01 & Text/(MCC, F0) & WaveNet \cite{conv_based_spoof}  \\ \hline
% A02 & Text/(MCC, F0, BAP) & WORLD \cite{world}\\ \hline
% A03 & Text/(MCC, F0, BAP) & WORLD \cite{world}\\ \hline
% A04 == A16 & Text/(MCC, F0) & Waveform Concatenation \\ \hline
% A05 & Speech^{\rm a}/(MFCC, F0, AP) & WORLD \cite{world}\\ \hline
% A06 == A19 & Speech^{\rm a}/LPC & Spectral filtering + OLA \\ \hline
% A07 & Text/(MCC, F0, BAP) & WORLD \cite{world} \\ \hline
% A08 & Text/(MCC, F0) & Neural source-filter \\ \hline
% A09 & Text/(MCC, F0) & Vocaine \cite{vocaine} \\ \hline
% A10 & Text/Mel-Spectrograms & WaveRNN \cite{wavernn} \\ \hline
% A11 & Text/Mel-Spectrograms & Griffin-Lim \cite{griffin} \\ \hline
% A12 & Text/(F0, Linguistic Features) & WaveNet \cite{conv_based_spoof} \\ \hline
% A13 & Speech^{\rm b}/MCC & Waveform filtering \\ \hline
% A14 & Speech^{\rm b}/(MCC, F0, BAP) & STRAIGHT \cite{straight} \\ \hline
% A15 & Speech^{\rm b}/(MCC, F0) & Waveform Concatenation \\ \hline
% A17 & Speech^{\rm a}/(MCC, F0) & Waveform filtering \\ \hline
% A18 & Speech^{\rm a}/MFCC & MFCC Vocoder \\ \hline
% Natural & - & - \\ \hline
% \end{tabular}
% \caption{ASVSpoof19 dataset description}
% \label{tab:asvspoof}
% \end{table*}

\begin{table}[h!]
\centering
\begin{tabular}{|p{1cm}|c|c|}
\hline
\textbf{Class} & \textbf{Input/Ouput} & \textbf{Generator} \\ \hline
A01 & Text/(MCC, F0) & WaveNet \cite{conv_based_spoof}  \\ \hline
A02 & Text/(MCC, F0, BAP) & WORLD \cite{world}\\ \hline
A03 & Text/(MCC, F0, BAP) & WORLD \cite{world}\\ \hline
A04, A16 & Text/(MCC, F0) & Waveform Concatenation \\ \hline
A05 & $Speech^{\rm a}$/(MFCC, F0, AP) & WORLD \cite{world}\\ \hline
A06, A19 & $Speech^{\rm a}$/LPC & Spectral filtering + OLA \\ \hline
A07 & Text/(MCC, F0, BAP) & WORLD \cite{world} \\ \hline
A08 & Text/(MCC, F0) & Neural source-filter \\ \hline
A09 & Text/(MCC, F0) & Vocaine \cite{vocaine} \\ \hline
A10 & Text/Mel-Spectrograms & WaveRNN \cite{wavernn} \\ \hline
A11 & Text/Mel-Spectrograms & Griffin-Lim \cite{griffin} \\ \hline
A12 & Text/(F0, Linguistic Features) & WaveNet \cite{conv_based_spoof} \\ \hline
A13 & $Speech^{\rm b}$/MCC & Waveform filtering \\ \hline
A14 & $Speech^{\rm b}$/(MCC, F0, BAP) & STRAIGHT \cite{straight} \\ \hline
A15 & $Speech^{\rm b}$/(MCC, F0) & Waveform Concatenation \\ \hline
A17 & $Speech^{\rm a}$/(MCC, F0) & Waveform filtering \\ \hline
A18 & $Speech^{\rm a}$/MFCC & MFCC Vocoder \\ \hline
Natural & - & - \\ \hline
\end{tabular}
\caption{ASVSpoof19 dataset description}
\label{tab:asvspoof}
\end{table}

\begin{comment}
\begin{table*}[]
\centering
\begin{tabular}{|c|c|c|c|}
\hline
\textbf{Class} & \textbf{ASVSpoof class} & \textbf{Input/Ouput} & \textbf{Generator} \\ \hline
1 & A01 & Text/(MCC, F0) & WaveNet \cite{conv_based_spoof}  \\ \hline
2 & A02 & Text/(MCC, F0, BAP) & WORLD \cite{world}\\ \hline
3 & A03 & Text/(MCC, F0, BAP) & WORLD \cite{world}\\ \hline
4 & A04 == A16 & Text/(MCC, F0) & Waveform Concatenation \\ \hline
5 & A05 & Speech^{\rm a}/(MFCC, F0, AP) & WORLD \cite{world}\\ \hline
6 & A06 == A19 & Speech^{\rm a}/LPC & Spectral filtering + OLA \\ \hline
7 & A07 & Text/(MCC, F0, BAP) & WORLD \cite{world} \\ \hline
8 & A08 & Text/(MCC, F0) & Neural source-filter \\ \hline
9 & A09 & Text/(MCC, F0) & Vocaine \cite{vocaine} \\ \hline
10 & A10 & Text/Mel-Spectrograms & WaveRNN \cite{wavernn} \\ \hline
11 & A11 & Text/Mel-Spectrograms & Griffin-Lim \cite{griffin} \\ \hline
12 & A12 & Text/(F0, Linguistic Features) & WaveNet \cite{conv_based_spoof} \\ \hline
13 & A13 & Speech^{\rm b}/MCC & Waveform filtering \\ \hline
14 & A14 & Speech^{\rm b}/(MCC, F0, BAP) & STRAIGHT \cite{straight} \\ \hline
15 & A15 & Speech^{\rm b}/(MCC, F0) & Waveform Concatenation \\ \hline
16 & A17 & Speech^{\rm a}/(MCC, F0) & Waveform filtering \\ \hline
17 & A18 & Speech^{\rm a}/MFCC & MFCC Vocoder \\ \hline
18 & Natural & - & - \\ \hline
\end{tabular}
\caption{ASVSpoof19 dataset description}
\label{tab:asvspoof}
\end{table*}
\end{comment}

\subsection{Dataset}
For conducting the studies, we utilize the \textit{Logical Access} portion of the data from the \textit{ASVSpoof 2019} data corpus\cite{asvspoof}. The description of each class is described in Table \ref{tab:asvspoof}. We have a total of 18 classes ($17$ Synthetic + $1$ Natural). However, the ASVSpoof19 dataset consists of distinct speakers in training and evaluation which best suits spoof classification (binary). Here the work is primarily concerned with synthetic speech algorithm classification; therefore we have pooled that data corpus and created custom data splits. We have created two versions of custom data splits as shown in Table \ref{tab:dataset},
\begin{itemize}
    \item Custom Split-1 or \textit{CS1}
    In this data split, the whole ASVSpoof19 data is accumulated and we have made 40\%, 10\%, and 50\% split for training, validation, and evaluation. We ensured that all the algorithms are uniformly spread over all three data sets.
    \item Custom Split-2 or \textit{CS2}
    To avoid overfitting the models by learning speaker information, we have created CS2 data splits such that there are distinct speakers in training and evaluation. We have 10 speakers in com
    mon for both training and validation, to ensure that model is performing relatively good on seen speakers. The proportions are mentioned in \ref{tab:dataset}.
\end{itemize}

\begin{table}[h!]
    \centering
    \caption{Dataset composition ($^{*}$ represents common speakers)}
    \begin{tabular}{|cc|c|c|}
    \hline
    \multicolumn{2}{|c|}{} & \textbf{CS1} & \textbf{CS2} \\ \hline
    \multicolumn{1}{|c|}{\multirow{3}{*}{\#Speakers}} & Train & 20 & 30 + 10* \\ \cline{2-4} 
    \multicolumn{1}{|c|}{} & Validation & 20 & 17 + 10* \\ \cline{2-4} 
    \multicolumn{1}{|c|}{} & Evaluation & 67 & 50 \\ \hline
    \multicolumn{1}{|c|}{\multirow{3}{*}{Proportion}} & Train & 40\% & 33.25\% \\ \cline{2-4} 
    \multicolumn{1}{|c|}{} & Validation & 10\% & 16.11\% \\ \cline{2-4} 
    \multicolumn{1}{|c|}{} & Evaluation & 50\% & 50.64\% \\ \hline
    \end{tabular}
    \label{tab:dataset}
\end{table}

In this work, we use speech signals synthesized at a sampling frequency of $16KHz$. To compute LP Residuals, we perform a $23rd$ order (for the reasons mentioned in section \ref{section:lpres_section}) LP analysis. To compute log mel spectrogram, we take $80$ mels, $512$ bin FFT, window length of $400$ samples ($25$ms), and hop length of $160$ ($10$ms).

\section{Results and Experiments}
In this section, evaluation performance and ablation analysis of the proposed architectures are studied. We train our models LPR+DNN and LMS+DNN and discuss the obtained performances. We perform various experiments on the proposed architectures. First, we fuse the networks with various fusing paradigms and validate the hypothesis of complementary information. Finally, we remove silence regions in the training data to prevent overfitting the data \cite{silence_artifacts}. This operation emphasizes on capturing the effects of phonemic variations in synthetic speech, rather than silent regions. 
\\

\textbf{Loss function}: We use cross entropy function as a loss function to optimize our DNN models, $\mathcal{L}_{(\boldsymbol{X}, \boldsymbol{y})}(\boldsymbol{\theta}) = H(\boldsymbol{y}, {LR_{1}}\left(f\left(\boldsymbol{\theta}, \boldsymbol{X}\right)\right))$. We use Adam optimizer to train our network.\\
\textbf{Evaluation Metrics}:- We report algorithm classification accuracy (in \%, higher the better) to evaluate the performance of our models.\\

\textbf{LPR-DNN vs LMS-DNN}: In Table \ref{tab:results}, we clearly observe that LPR-DNN is outperforming LMS-DNN. This is evident from the hypothesis we gave earlier. LP Residuals have more discriminative features because we are removing the VTS response by inverse filtering it because the speech generation algorithms typically optimize 2nd order features. The best performance obtained among these two algorithms is by LPR-DNN with an algorithm classification accuracy of 98.95\%.

\begin{table}[h!]
\centering
\caption{Results of various models (Algorithm Accuracy \%)}
\resizebox{\columnwidth}{!}{%
\begin{tabular}{|c|c|c|c|}
\hline
\multirow{2}{*}{\textbf{Classifier type}} & \multirow{2}{*}{\textbf{Architecture}} & \textbf{CS1} & \textbf{CS2} \\ \cline{3-4} 
 &  & \textbf{Evaluation} & \textbf{Evaluation} \\ \hline
\multirow{4}{*}{Without silence removal} & LPR-DNN & 98.95\%&  96.02\%\\ \cline{2-4} 
 & LMS-DNN & 96.42\% &  96.04\%\\ \cline{2-4} 
 & LPR+LMS-DNN\textsuperscript{*} & \textbf{99.58\%} & 
 \textbf{98.66\%}\\ \cline{2-4} 
 & LMS-LPR-DNN & 99.06\% &  {97.65\%} \\ \hline \hline
\multirow{4}{*}{With silence removal} & LPR-DNN & 96.37\% & 98.42\% \\ \cline{2-4} 
 & LMS-DNN & 83.08\% & 90.44\% \\ \cline{2-4} 
 & LPR+LMS-DNN\textsuperscript{*} & \textbf{98.66\%} & \textbf{99.27\%} \\ \cline{2-4} 
 & LPR+LMS-DNN & {98.57\%} & {96.29\%} \\ \hline
\end{tabular}
}
\label{tab:results}
\end{table}

\begin{figure*}[ht!]
    \includegraphics[width=2\columnwidth, height=5cm]{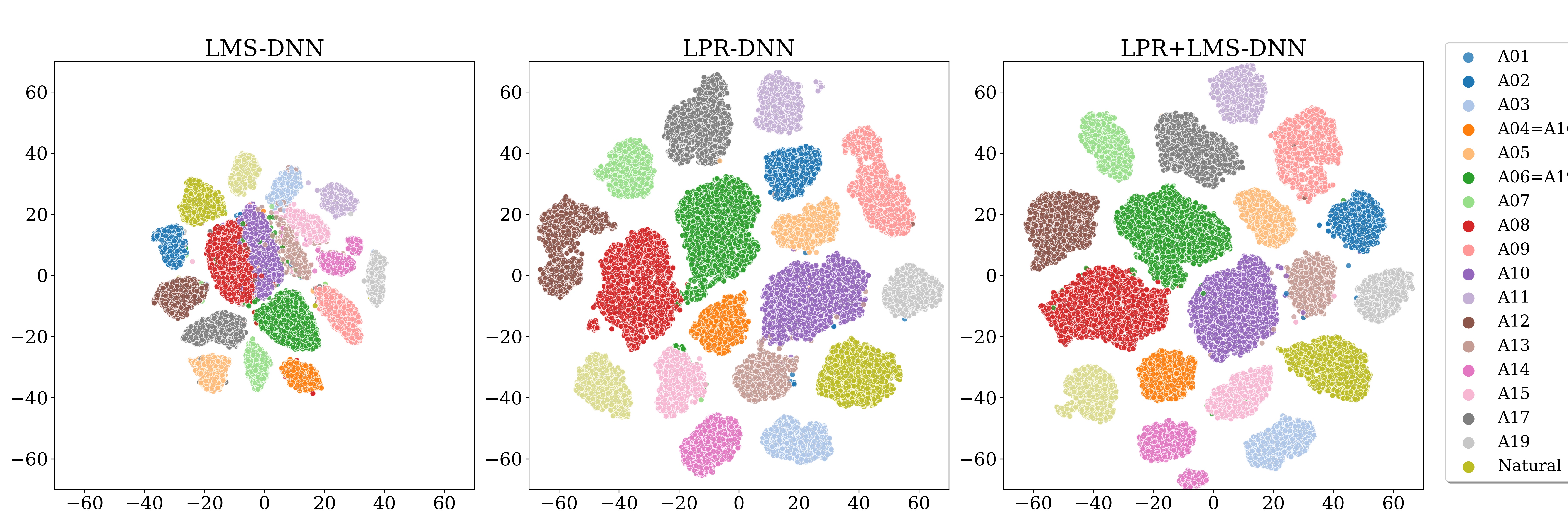}
    \caption{t-SNE embeddings for processed vector from X-vectors}
    \label{fig:tsne}
\end{figure*}

\subsection{Model Fusion}
Employing various fusion paradigms in speech community \cite{fusion_2, fusion_1} 
 have shown significant performance leaps. As discussed earlier, VS and VTS features have complementary information, which we would like to exploit in this experiment by performing two different kinds of fusion.

\subsubsection{LPR+LMS-DNN: Intermediate fusion \cite{ramachandram_sp2017}}
This fusion paradigm is achieved by fusing features at the architecture level; typically, the concatenation of features occurs in the common latent space of different systems. In this architecture variant, we concatenate features obtained after frame-level encoding and perform multi-headed self-attention, statistical pooling, and logistic regression. 

\subsubsection{LPR+LMS-DNN\textsuperscript{*}: Late fusion \cite{ramachandram_sp2017}}
One elementary fusion technique is known as score level fusion, where we perform weighted averages on logits obtained from different systems. We perform the same using LPR-DNN and LMS-DNN with equal weights.

\subsection{Comparisons among variants of proposed models}

The proposed models are hyper-tuned to achieve the best performance, and Table \ref{tab:results} consists of the performances achieved. To avoid over-fitting DNNs due to distinct silent regions present in spoof algorithms\cite{silence_artifacts}, we also experimented with silence removal using the Voice Activity Detector (VAD) system \cite{yang2021torchaudio}. Below are the four major points that we would like to accentuate.
\subsubsection{VS features are a cut above VTS features}
As we observed discriminative features of VS in the previous sections, it is evident from the performance on LPR-DNN, which is at least $2\%$ superior to LMS-DNN. 

\subsubsection{Performance leap with fusion techniques}
As anticipated, fusion techniques showed better performance than LPR-DNN and LMS-DNN, which is a popular paradigm in the DNN community to achieve better results. This performance leap is observed in all the studies performed, such as with/without silence removal and inference with distinct speakers. We plotted t-SNE \cite{tsne} embeddings onto $2D$ space to evaluate the separability, which is shown in Figure \ref{fig:tsne}. It is observed that clusters in LMS-DNN embeddings are merged into others and are close together. However, after performing feature fusion, the clusters are clearly separated.
This is yet another piece of evidence for the hypothesis that VS and VTS features consist of complementary information.

\subsubsection{Over-fitting of VTS features}
On the \textit{CS1} data set, the performance dip when silence regions are removed in LMS-DNN is significant, implying that LMS-DNN is over-fitting the data by exploiting the amount of silence region present in synthetic speech signals. In contrast, LPR-DNN shows robustness to the higher amount of silence regions.

\subsubsection{Robustness with unseen speakers}
Speaker over-fitting is a generic issue in speech recognition/verification. To test this, we have created \textit{CS2} dataset and trained on it appropriately. The results obtained have very slight variations compared to \textit{CS1}, implying that proposed DNNs are robust to new speakers and can identify the underlying algorithm characteristics.

\section{Analysis and Inference}
\label{subsection : highlighted_seg}
In this section, we analyze the self-attentive modules and visualize the portion of the speech signal highlighted. We pass a speech signal with an utterance of "\textit{THERE WAS A SUBSTANTIAL EXPLOSION}" to both, LP residual and Mel Spectrogram-based Models. We look at the attention values, which is a matrix of size $(\#Channels, T')$, and $T'$ is determined by the effective receptive field and stride as mentioned in \ref{tab:receptive_field}. For visualization, we take the mean of the matrix along the channel axis and later binarize it with the mean over time as the threshold.

From Figure \ref{fig:atten}, we find that LPR-DNN gives more attention to the transition of phonemes. In literature, it is known as \textit{coarticulation}\cite{coarticulation}. This is evident from formant plots where the green bands(LPR-DNN) are highlighted at the portions where formants change significantly, indicating the change in phoneme. In contrast, LMS-DNN focuses on stationary parts of phonemes where the formants are relatively constant.

Typically, when there is a change from voiced-to-unvoiced sound or vice-versa, the source signal changes from random noise to impulse trains. This change is captured by LPR-DNN, and it is different for different algorithms, therefore providing stronger discrimination compared to LMS-DNN. 
\begin{figure}[h]
    \centering
    \includegraphics[width=\columnwidth]{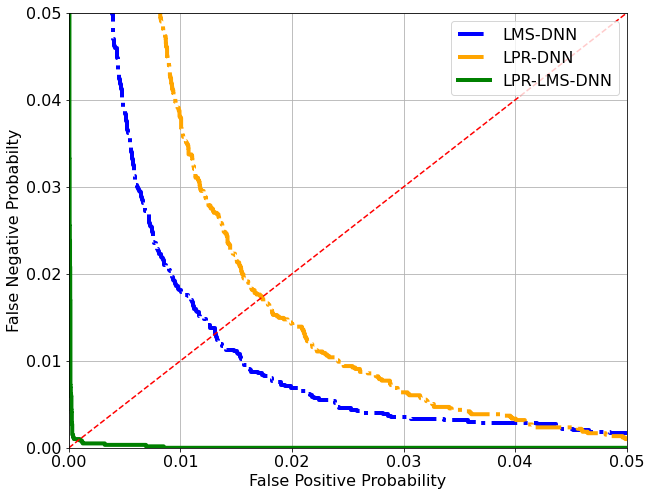}
    \caption{ROC plots for finding EER of LMS-DNN, LPR-DNN and fused LPR-LMS-DNN. The EER for the case of LPR-LMS-DNN (green) is observed to be so low that the plot almost overlaps with the axes }
    \label{fig:eer}
\end{figure}

To observe the benefit of utilizing the complementariness of the features captured by LPR-DNN and LMS-DNN, we take the case of the binary spoof classifier. By plotting the ROC curve, we find that there is a significant reduction of EER value for the fused model when compared to the individual models. For LPR-DNN and LMS-DNN, the EER is 0.013114 and 0.017282 respectively, whereas the EER for the fused model is 0.000989. By maximizing the information extracted from the speech signal, we can observe an increase in classification accuracy.

\begin{figure*}[h!]
    \centering
    \begin{subfigure}[b]{ \columnwidth}
        \centering
        \includegraphics[width=\columnwidth,height=10cm,trim={0 0 {0.6\columnwidth} 0},clip]{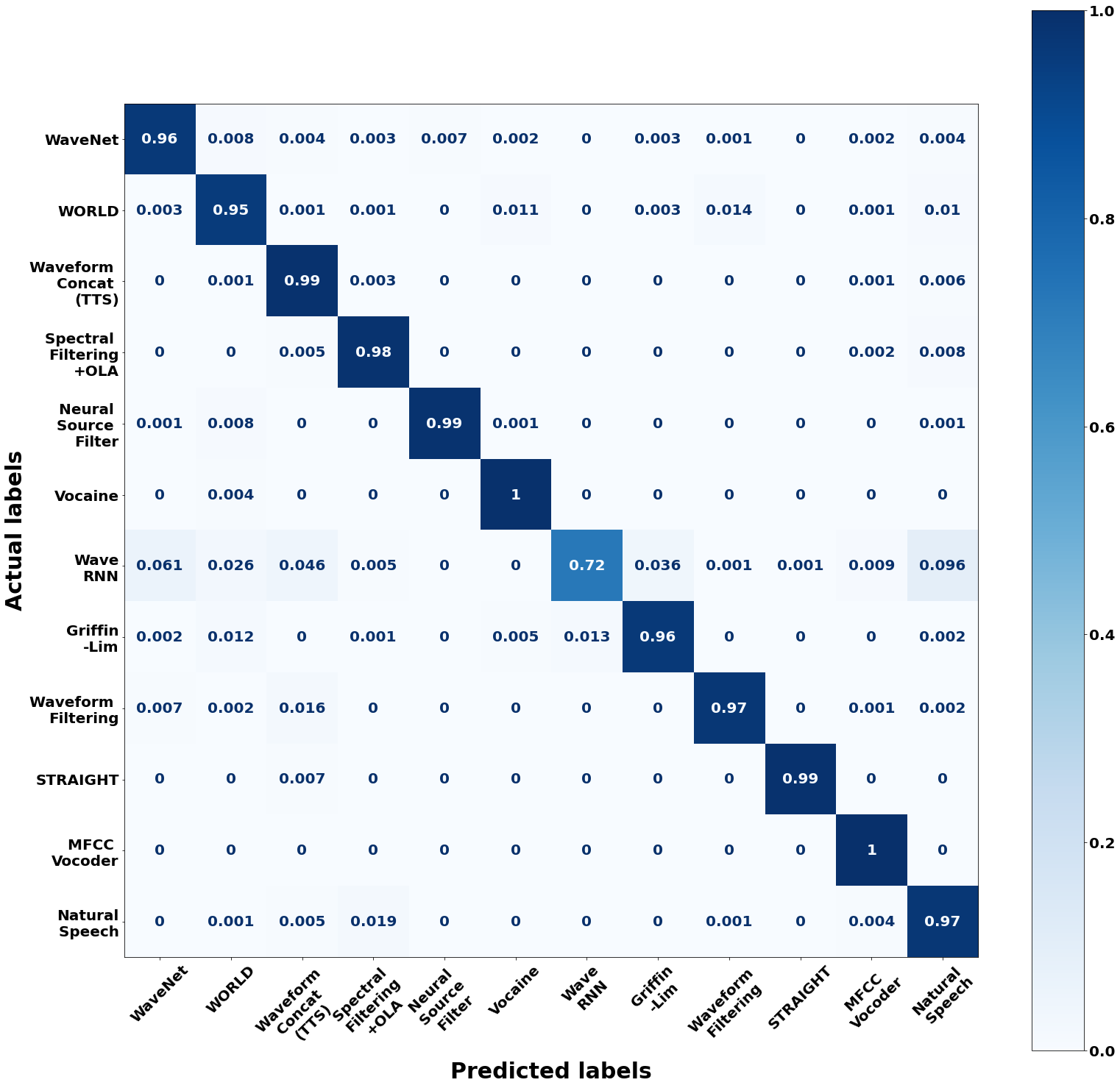}
        \caption{LMS-DNN}
        \label{fig:cm_lms}
    \end{subfigure}
    % \hspace{0.1cm}
    \begin{subfigure}[b]{ \columnwidth}
        \centering
        \includegraphics[width= 1.1\columnwidth,height=10cm]{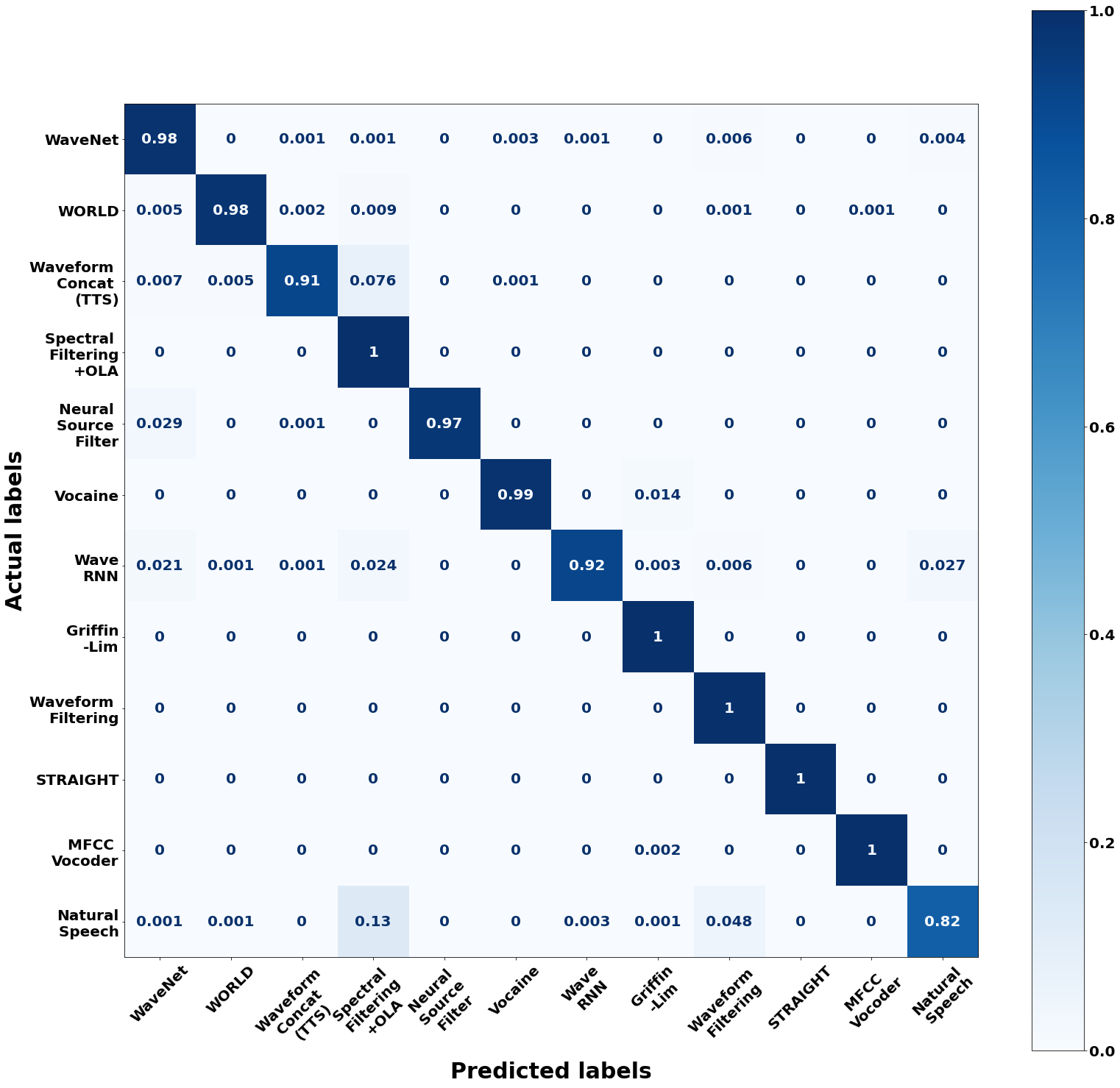}
        \caption{LPR-DNN}
        \label{fig:cm_lpres}
    \end{subfigure}
    \label{fig:cm_fusion}
    \caption{Confusion matrix plots for proposed networks based on the waveform generator of the synthesizer}
    \label{fig:cm_plots}
\end{figure*} 

To understand the attributions of the spoof samples, we have analyzed the confusion plots based on the waveform generator of the underlying synthesizer as shown in Figure \ref{fig:cm_plots}. The confusion plots suggest that the synthesizers are identified correctly for most of the waveform generators. It can be observed from Figure \ref{fig:cm_lms} that the LMS-DNN performs better on waveform generators which are based on Waveform Concatenation, Neural Source Filter, and Vocaine.

On the contrary, Figure \ref{fig:cm_lpres} shows that the rest of the synthesizers are much more accurately identified by the LPR-DNN. It can be observed that the LPR-DNN performance is much more pronounced in the case of Griffin-Lim, Waveform Filtering, STRAIGHT and Spectral Filtering + OLA. {However, we observe that the spoof samples generated from the WaveRNN seem to be the most confusing class for both the designed architectures. The confusion happens to be between WaveRNN and the speech samples generated by Wavenet and the natural class.}

\section{Conclusion}
{We proposed a spoof speech algorithm classifier using DNN by using voice source (LP Residuals) and vocal tract (Mel Spectrograms) features. By combining evidence from these two features, we achieved a performance leap indicating complementary information between them. In a further investigation into the attention maps of DNN over the spoof speech, we observed that LPR-DNN focused on phoneme transitions, and in contrast, LMS-DNN is focused more on the stationary portion of phonemes. We conclude that artifacts in spoof speech are deductive using higher-order correlations and the research in spoof speech generation algorithms has to consider improving natural phonetic transitions. This study could be utilized for enhancing the synthetic generation techniques by improving on the defects identified. Also, an exhaustive study on phonemes could be performed to identify which phoneme utterance could be more critical in spoof detection.}

\section{Acknowledgement}
    We sincerely thank SIP Lab, IIT Hyderabad, for providing the computation power needed for the studies and the IEEE SP Cup 2022 organizers for the problem statement and the initial dataset. 
    % TUKR and KPK would like to acknowledge the \textit{Interstellar} soundtrack for inspiring the writing.

\bibliographystyle{ieeetr}
\bibliography{refs}

\begin{thebibliography}{10}

\bibitem{fraudcase_1}
C.~Stupp, ``Fraudsters {Used} {AI} to {Mimic} {CEO}\textquoteright{}s {Voice}
  in {Unusual} {Cybercrime} {Case}.''
  https://www.wsj.com/articles/fraudsters-use-ai-to-mimic-ceos-voice-in-unusual-cybercrime-case-11567157402,
  Aug 30 2019.

\bibitem{fraudcase_2}
D.-S.-S. Engineering-Team, ``When acting turns criminal: Deepfakes and voice
  impersonators in the cybercriminal underground.''
  https://www.digitalshadows.com/blog-and-research/deepfakes-and-voice-impersonators-in-the-cybercriminal-underground/,
  Dec 1 2021.

\bibitem{high_freq_artifacts}
J.~Yang and R.~K. Das, ``Long-term high frequency features for synthetic speech
  detection,'' {\em Digital Signal Processing}, vol.~97, p.~102622, 2020.

\bibitem{feature_comp_synth_detect}
M.~Sahidullah, T.~Kinnunen, and C.~Hanilçi, ``A comparison of features for
  synthetic speech detection,'' 09 2015.

\bibitem{dnn_deepfake_detect}
H.~Dhamyal, A.~Ali, I.~A. Qazi, and A.~A. Raza, ``Using self attention dnns to
  discover phonemic features for audio deep fake detection,'' in {\em 2021 IEEE
  Automatic Speech Recognition and Understanding Workshop (ASRU)},
  pp.~1178--1184, 2021.

\bibitem{silence_artifacts}
D.~Mari, F.~Latora, and S.~Milani, ``The sound of silence: Efficiency of first
  digit features in synthetic audio detection,'' 2022.

\bibitem{lp_speaker_recog_1}
H.~Wakita, ``Residual energy of linear prediction applied to vowel and speaker
  recognition,'' {\em IEEE Transactions on Acoustics, Speech, and Signal
  Processing}, vol.~24, no.~3, pp.~270--271, 1976.

\bibitem{lp_speaker_recog_2}
M.~Faundez-Zanuy and D.~Rodriguez-Porcheron, ``Speaker recognition using
  residual signal of linear and nonlinear prediction models,'' 01 1998.

\bibitem{lp_speaker_recog_3}
W.-C. Hsu and J.-N. Sun, ``A study of the usefulness of linear prediction
  residual for speaker recognition,'' {\em Advanced Science Letters}, vol.~9,
  pp.~754--761, 04 2012.

\bibitem{lp_speaker_ver}
W.-C. Hsu, W.-H. Lai, and W.-P. Hong, ``Usefulness of residual-based features
  in speaker verification and their combination way with linear prediction
  coefficients,'' in {\em Ninth IEEE International Symposium on Multimedia
  Workshops (ISMW 2007)}, pp.~246--251, 2007.

\bibitem{lp_speaker_info}
S.~{Mahadeva Prasanna}, C.~S. Gupta, and B.~Yegnanarayana, ``Extraction of
  speaker-specific excitation information from linear prediction residual of
  speech,'' {\em Speech Communication}, vol.~48, no.~10, pp.~1243--1261, 2006.

\bibitem{xvect}
D.~Snyder, D.~Garcia-Romero, G.~Sell, D.~Povey, and S.~Khudanpur, ``X-vectors:
  Robust dnn embeddings for speaker recognition,'' in {\em 2018 IEEE
  International Conference on Acoustics, Speech and Signal Processing
  (ICASSP)}, pp.~5329--5333, 2018.

\bibitem{speech_prod_complete}
A.~Lacroix, ``Speech production-physics, models and prospective applications,''
  in {\em ISPA 2001. Proceedings of the 2nd International Symposium on Image
  and Signal Processing and Analysis. In conjunction with 23rd International
  Conference on Information Technology Interfaces (IEEE Cat.}, pp.~3--, 2001.

\bibitem{arx_model}
W.~Zhu and H.~Kasuya, ``A new speech synthesis system based on the arx speech
  production model,'' in {\em Proceeding of Fourth International Conference on
  Spoken Language Processing. ICSLP '96}, vol.~3, pp.~1413--1416 vol.3, 1996.

\bibitem{dynamic_model}
K.~Iso, ``Speech recognition using dynamical model of speech production,'' in
  {\em 1993 IEEE International Conference on Acoustics, Speech, and Signal
  Processing}, vol.~2, pp.~283--286 vol.2, 1993.

\bibitem{GA_model}
L.~Mitiche and A.~B.~H. Adamou-Mitiche, ``Second order speech model based on
  ga's,'' in {\em 2016 4th International Conference on Control Engineering \&
  Information Technology (CEIT)}, pp.~1--4, 2016.

\bibitem{models_survey}
C.-W. Kim, ``Models of speech production,'' in {\em Formal Aspects of Cognitive
  Processes} (T.~Storer and D.~Winter, eds.), (Berlin, Heidelberg),
  pp.~142--158, Springer Berlin Heidelberg, 1975.

\bibitem{source_filter}
G.~Fant, {\em Acoustic theory of speech production}.
\newblock No.~2, Walter de Gruyter, 1970.

\bibitem{anthropometry_voice}
R.~Singh, B.~Raj, and D.~Gencaga, ``Forensic anthropometry from voice: An
  articulatory-phonetic approach,'' in {\em 2016 39th International Convention
  on Information and Communication Technology, Electronics and Microelectronics
  (MIPRO)}, pp.~1375--1380, 2016.

\bibitem{jitter_spoof}
A.~J. Rozsypal and B.~F. Millar, ``Perception of jitter and shimmer in
  synthetic vowels,'' {\em Journal of Phonetics}, vol.~7, no.~4, pp.~343--355,
  1979.

\bibitem{coarticulation}
P.~Menzerath and A.~de~Carvalho~Lacerda, ``Koartikulation, steuerung und
  lautabgrenzung : eine experimentelle untersuchung,'' 1933.

\bibitem{story_2017}
B.~H. Story and K.~Bunton, ``An acoustically-driven vocal tract model for stop
  consonant production,'' {\em Speech Communication}, vol.~87, pp.~1--17, 2017.

\bibitem{birkholz_2013}
P.~Birkholz, ``Modeling consonant-vowel coarticulation for articulatory speech
  synthesis,'' {\em PLOS ONE}, vol.~8, pp.~1--17, 04 2013.

\bibitem{MFCC_alzheimers}
A.~Meghanani, A.~C.~S., and A.~G. Ramakrishnan, ``An exploration of log-mel
  spectrogram and mfcc features for alzheimer’s dementia recognition from
  spontaneous speech,'' in {\em 2021 IEEE Spoken Language Technology Workshop
  (SLT)}, pp.~670--677, 2021.

\bibitem{env_sound_classify}
F.~Beritelli and R.~Grasso, ``A pattern recognition system for environmental
  sound classification based on mfccs and neural networks,'' in {\em 2008 2nd
  International Conference on Signal Processing and Communication Systems},
  pp.~1--4, 2008.

\bibitem{poser_1990}
W.~J. Poser, ``Douglas o'shaughnessy, speech communication: Human and machine.
  reading, massachusetts: Addison-wesley publishing company, 1987. pp. xviii
  568. isbn 0-201-16520-1.,'' vol.~20, p.~52–54, Cambridge University Press,
  1990.

\bibitem{acoustic_event_detect}
C.~V. Cotton and D.~P.~W. Ellis, ``Spectral vs. spectro-temporal features for
  acoustic event detection,'' in {\em 2011 IEEE Workshop on Applications of
  Signal Processing to Audio and Acoustics (WASPAA)}, pp.~69--72, 2011.

\bibitem{auditory_scene_recog}
V.~Peltonen, J.~Tuomi, A.~Klapuri, J.~Huopaniemi, and T.~Sorsa, ``Computational
  auditory scene recognition,'' in {\em 2002 IEEE International Conference on
  Acoustics, Speech, and Signal Processing}, vol.~2, pp.~II--1941--II--1944,
  2002.

\bibitem{logmel_asfeature}
Y.~Gao, T.~Vuong, M.~Elyasi, G.~Bharaj, and R.~Singh, ``Generalized spoofing
  detection inspired from audio generation artifacts,'' 2021.

\bibitem{receptive_field}
W.~Luo, Y.~Li, R.~Urtasun, and R.~Zemel, ``Understanding the effective
  receptive field in deep convolutional neural networks,'' 2017.

\bibitem{spkrverif}
F.~Bimbot, J.-F. Bonastre, C.~Fredouille, G.~Gravier, I.~Magrin-Chagnolleau,
  S.~Meignier, T.~Merlin, J.~Ortega-Garc{\'\i}a, D.~Petrovska-Delacr{\'e}taz,
  and D.~A. Reynolds, ``A tutorial on text-independent speaker verification,''
  {\em EURASIP Journal on Advances in Signal Processing}, vol.~2004, no.~4,
  pp.~1--22, 2004.

\bibitem{tachetron}
Y.~Wang, R.~Skerry-Ryan, D.~Stanton, Y.~Wu, R.~J. Weiss, N.~Jaitly, Z.~Yang,
  Y.~Xiao, Z.~Chen, S.~Bengio, Q.~Le, Y.~Agiomyrgiannakis, R.~Clark, and R.~A.
  Saurous, ``Tacotron: Towards end-to-end speech synthesis,'' 2017.

\bibitem{wavenet}
A.~v.~d. Oord, S.~Dieleman, H.~Zen, K.~Simonyan, O.~Vinyals, A.~Graves,
  N.~Kalchbrenner, A.~Senior, and K.~Kavukcuoglu, ``Wavenet: A generative model
  for raw audio,'' 2016.

\bibitem{deep_voice}
S.~O. Arik, M.~Chrzanowski, A.~Coates, G.~Diamos, A.~Gibiansky, Y.~Kang, X.~Li,
  J.~Miller, A.~Ng, J.~Raiman, S.~Sengupta, and M.~Shoeybi, ``Deep voice:
  Real-time neural text-to-speech,'' 2017.

\bibitem{evidence_res_ph_mfcc}
K.~Murty and B.~Yegnanarayana, ``Combining evidence from residual phase and
  mfcc features for speaker recognition,'' vol.~13, pp.~52--55, 2006.

\bibitem{better_mfcc_feature}
R.~Gonzalez, {\em Better Than MFCC Audio Classification Features},
  pp.~291--301.
\newblock 10 2013.

\bibitem{synth_uttr_roman}
G.~Pop and D.~Burileanu, ``Towards detection of synthetic utterances in
  romanian language speech forensics,'' in {\em 2021 International Conference
  on Speech Technology and Human-Computer Dialogue (SpeD)}, pp.~80--84, 2021.

\bibitem{allpass_modelling_lpr}
K.~S.~R. Murty, V.~Boominathan, and K.~Vijayan, ``Allpass modeling of lp
  residual for speaker recognition,'' in {\em 2012 International Conference on
  Signal Processing and Communications (SPCOM)}, pp.~1--5, 2012.

\bibitem{LP_tutorial}
J.~Makhoul, ``Linear prediction: A tutorial review,'' {\em Proceedings of the
  IEEE}, vol.~63, no.~4, pp.~561--580, 1975.

\bibitem{rabiner_book}
L.~R. Rabiner, B.~Gold, and C.~K. Yuen, ``Theory and application of digital
  signal processing,'' {\em IEEE Transactions on Systems, Man, and
  Cybernetics}, vol.~8, no.~2, pp.~146--146, 1978.

\bibitem{quatieri_book}
T.~Quatieri, {\em Discrete-time Speech Signal Processing: Principles and
  Practice}.
\newblock Pearson Education, 2002.

\bibitem{cepstral_bispectral_analysis}
A.~K. Singh and P.~Singh, ``Detection of ai-synthesized speech using cepstral
  \& bispectral statistics,'' in {\em 2021 IEEE 4th International Conference on
  Multimedia Information Processing and Retrieval (MIPR)}, pp.~412--417, 2021.

\bibitem{bicoher_phase}
C.~K. Kovach, H.~Oya, and H.~Kawasaki, ``The bispectrum and its relationship to
  phase-amplitude coupling,'' {\em NeuroImage}, vol.~173, pp.~518--539, 2018.

\bibitem{dropout}
N.~Srivastava, G.~Hinton, A.~Krizhevsky, I.~Sutskever, and R.~Salakhutdinov,
  ``Dropout: A simple way to prevent neural networks from overfitting,'' {\em
  Journal of Machine Learning Research}, vol.~15, no.~56, pp.~1929--1958, 2014.

\bibitem{layer_norm}
J.~L. Ba, J.~R. Kiros, and G.~E. Hinton, ``Layer normalization,'' 2016.

\bibitem{conv_based_spoof}
A.~van~den Oord, S.~Dieleman, H.~Zen, K.~Simonyan, O.~Vinyals, A.~Graves,
  N.~Kalchbrenner, A.~Senior, and K.~Kavukcuoglu, ``Wavenet: A generative model
  for raw audio,'' in {\em Arxiv}, 2016.

\bibitem{world}
M.~MORISE, F.~YOKOMORI, and K.~OZAWA, ``World: A vocoder-based high-quality
  speech synthesis system for real-time applications,'' {\em IEICE Transactions
  on Information and Systems}, vol.~E99.D, no.~7, pp.~1877--1884, 2016.

\bibitem{vocaine}
Y.~Agiomyrgiannakis, ``Vocaine: The vocoder and applications in speech
  synthesis,'' pp.~4230--4234, 04 2015.

\bibitem{wavernn}
N.~Kalchbrenner, E.~Elsen, K.~Simonyan, S.~Noury, N.~Casagrande, E.~Lockhart,
  F.~Stimberg, A.~v.~d. Oord, S.~Dieleman, and K.~Kavukcuoglu, ``Efficient
  neural audio synthesis,'' 2018.

\bibitem{griffin}
D.~Griffin and J.~Lim, ``Signal estimation from modified short-time fourier
  transform,'' {\em IEEE Transactions on Acoustics, Speech, and Signal
  Processing}, vol.~32, no.~2, pp.~236--243, 1984.

\bibitem{straight}
H.~Kawahara, ``Straight, exploitation of the other aspect of vocoder:
  Perceptually isomorphic decomposition of speech sounds,'' {\em Acoustical
  Science and Technology}, vol.~27, no.~6, pp.~349--353, 2006.

\bibitem{asvspoof}
X.~Wang, J.~Yamagishi, M.~Todisco, H.~Delgado, A.~Nautsch, N.~Evans,
  M.~Sahidullah, V.~Vestman, T.~Kinnunen, K.~A. Lee, L.~Juvela, P.~Alku, Y.-H.
  Peng, H.-T. Hwang, Y.~Tsao, H.-M. Wang, S.~L. Maguer, M.~Becker,
  F.~Henderson, R.~Clark, Y.~Zhang, Q.~Wang, Y.~Jia, K.~Onuma, K.~Mushika,
  T.~Kaneda, Y.~Jiang, L.-J. Liu, Y.-C. Wu, W.-C. Huang, T.~Toda, K.~Tanaka,
  H.~Kameoka, I.~Steiner, D.~Matrouf, J.-F. Bonastre, A.~Govender, S.~Ronanki,
  J.-X. Zhang, and Z.-H. Ling, ``Asvspoof 2019: A large-scale public database
  of synthesized, converted and replayed speech,'' 2019.

\bibitem{fusion_2}
N.~Chauhan, T.~Isshiki, and D.~Li, ``Speaker recognition using fusion of
  features with feedforward artificial neural network and support vector
  machine,'' in {\em 2020 International Conference on Intelligent Engineering
  and Management (ICIEM)}, pp.~170--176, 2020.

\bibitem{fusion_1}
S.~Toshniwal, A.~Kannan, C.-C. Chiu, Y.~Wu, T.~N. Sainath, and K.~Livescu, ``A
  comparison of techniques for language model integration in encoder-decoder
  speech recognition,'' 2018.

\bibitem{ramachandram_sp2017}
D.~Ramachandram and G.~W. Taylor, ``Deep multimodal learning: A survey on
  recent advances and trends,'' {\em IEEE Signal Processing Magazine}, vol.~34,
  no.~6, pp.~96--108, 2017.

\bibitem{yang2021torchaudio}
Y.-Y. Yang, M.~Hira, Z.~Ni, A.~Chourdia, A.~Astafurov, C.~Chen, C.-F. Yeh,
  C.~Puhrsch, D.~Pollack, D.~Genzel, D.~Greenberg, E.~Z. Yang, J.~Lian,
  J.~Mahadeokar, J.~Hwang, J.~Chen, P.~Goldsborough, P.~Roy, S.~Narenthiran,
  S.~Watanabe, S.~Chintala, V.~Quenneville-Bélair, and Y.~Shi, ``Torchaudio:
  Building blocks for audio and speech processing,'' {\em arXiv preprint
  arXiv:2110.15018}, 2021.

\bibitem{tsne}
L.~van~der Maaten and G.~Hinton, ``Visualizing data using t-sne,'' {\em Journal
  of Machine Learning Research}, vol.~9, no.~86, pp.~2579--2605, 2008.

\end{thebibliography}

\end{document}